# Thermally Desorbable InN Capping Layers for Nitride Surface Science

Mellie Lemon,[1] Amitayush Thakur,[2] Anthony Rice,[1] Glenn Teeter,[1] Michelle Smeaton,[1] Renae Gannon,[1] Jessica L. McChesney,[3] M. Brooks Tellekamp[1]

[1]*National Laboratory of the Rockies, Golden, CO, USA*
[2]*Advanced Photon Source, Argonne National Laboratory, Lemont, IL, USA*
[3]*Materials Science Division, Argonne National Laboratory, Lemont, IL, USA*

Group III-nitride thin films are essential for optoelectronic and power devices, where surface and interface quality critically influence performance. It is often necessary to transfer these films through atmosphere for processing or characterization steps, which can introduce significant surface contamination. Here we demonstrate a technique to protect the surface of III-N films during atmospheric exposure by capping with sacrificial InN layers. Using molecular beam epitaxy grown AlGaN films as our representative protected material, we take advantage of the lower decomposition and desorption temperatures of InN and metallic In to remove the protective cap layer *in situ* via thermal desorption without damaging the AlGaN film beneath. Angle-resolved photoemission spectroscopy (ARPES) measurements of the valence band dispersion in $Al_{0.4}Ga_{0.6}N$ demonstrate that the film surface is recovered after InN decapping, highlighting the versatility of this process in allowing further surface-sensitive characterization of films that had been exposed to air.

## 1. INTRODUCTION

There is broad research interest in group III-nitride materials due to their useful properties making them suitable for applications such as high frequency communication,[1] LEDs,[2] and power electronics.[3,4] Due to a lack of low-cost native substrates, III-nitride compounds are typically heteroepitaxially integrated with other materials. The termination, polarity, bonding reconstructions, and defects at the surface of III-nitrides can significantly impact interfacial properties such as electron and phonon transport. For example, a two-dimensional electron gas (2DEG) forms at the GaN/AlGaN interface.[5] Despite being discovered over 30 years prior, the high oxygen affinity of III-nitrides, particularly AlN and AlGaN, has meant that electron structure measurements like photoemission were limited to *in situ* grown films.[6–9] Only recently, angle-resolved photoelectron spectroscopy (ARPES) was performed on AlGaN/GaN structures using soft X-ray synchrotron radiation to access the 2DEG at the buried interface, leading to the discovery of planar anisotropy of the electron Fermi surface at the interface, which impacts transport properties.[10] This discovery gave insight into how to manipulate GaN/AlN/AlGaN interfaces to optimize properties for both high power and high frequency transistors.

Further research into the surface science of III-N thin films will lead to valuable insights into surface and interfacial electron and phonon transport. However, research is limited by the inability to probe non-pristine film surfaces because surface-sensitive measurements can easily become saturated with unwanted signal from contamination and defects. It is frequently necessary to transport thin films through atmosphere for post-growth characterization and processing, but atmospheric exposure can lead to significant changes at the film surface. Exposure to air can prompt oxide formation or contamination via adsorption of adventitious carbon or cyclic siloxanes, with the latter proving particularly disadvantageous due to Si-doping effects.[11–17] While contaminants can be removed through surface treatment methods like solvent cleaning or ion sputtering, this can lead to adsorption of residual solvent species or to damage from ion bombardment.[18–20] An alternative method to reduce contamination during air transfer is the use of a protective cap layer. In the well-studied III-V (V = As, P, Sb) system, it has been demonstrated that capping with an amorphous As layer protects the film surface during transport in air and can be completely removed by thermal desorption.[21–24] Based on this work, we were motivated to seek out a reversible, protective capping material for III-N thin films.

Since the group III metals form highly stable oxide compounds, amorphous Al or Ga layers were not practical options for capping. Amorphous In has been demonstrated as a capping layer for GaN/AlN,[25] but the tendency to form $In_2O_3$ in atmosphere can lead to a residual partial pressure of oxygen during the decapping process that could oxidize surface species. Among the binary III-N materials, InN has a relatively low decomposition temperature of 435°C[26] vs 800-900°C for GaN[27,28] and 1800°C for AlN.[29] Additionally, InN forms only weakly-bonded, amorphous indium oxynitride phases in atmosphere, reducing the potential for re-introduction of oxygen to the system during thermal removal.[30,31] Due to its promising qualities, we elected to investigate InN as a sacrificial, protective cap layer to enable III-N surface studies.

We report the deposition of InN caps on AlGaN films grown by molecular beam epitaxy (MBE). We demonstrate the reversibility of the InN protective layer by *in situ* thermal desorption at temperatures below the decomposition limit for AlGaN and show that the AlGaN surface is unaffected by the decapping process. The necessity for a protective layer is emphasized by x-ray photoelectron spectroscopy (XPS) measurements of surface oxygen and carbon concentrations for protected vs nonprotected films. The advantages of using an InN protective cap layer for surface-sensitive studies are highlighted through ARPES measurements of the band dispersion and isoenergy surface of an AlGaN film enabled by *in situ* cap removal in vacuum after exposing the film stack to atmosphere for transport. Though AlGaN was used as a model system, InN capping layers show promise for protection of a wide variety of MBE-grown compounds, including AlN alloy ferroelectrics (AlScN,[32] AlBN,[33] AlYN,[34] AlLaN,[35] etc.)[36] and transition metal nitrides.[37,38] The use of this capping layer will facilitate surface studies to unveil phenomena that would otherwise be obscured by contaminant effects or by damage from oxide removal through sputtering.

## 2. EXPERIMENTAL METHODS

### Thin film growth and characterization

Films were grown epitaxially on 10x10 mm (0001) oriented AlN templates on c-plane sapphire substrates provided by Kyma Technologies. The substrates were backside metallized with 1 μm of tantalum to ensure even heating across the area during growth. The substrates were cleaned by sequentially soaking in acetone, methanol, then isopropyl alcohol with deionized water rinses between each solvent step. Plasma-assisted molecular beam epitaxy (PA-MBE) was carried out in a Riber Compact 21T MBE system. Cleaned substrates were loaded into Mo platens and outgassed in an introductory vacuum chamber at 150°C for at least 2 hours before transferring to an intermediate buffer chamber. The substrate temperature during film growth was controlled using a radiative SiC heater, and the temperature was monitored by a thermocouple that was calibrated using temperature and flux-dependent gallium desorption experiments.

The Ga was sourced from a SUMO style effusion cell while In and Al were sourced from conventional dual-filament effusion cells. The sourced material purity was 7N for Ga, 6N for Al, and 6N5 for In. The Ga, Al, and In evaporation rates were measured using a retractable beam flux monitor prior to growth. A reactive nitrogen plasma was supplied during growth through a 13.56 MHz radio-frequency source (Veeco UNIBulb) operating at 350W and a flow of 2.6 SCCM for AlN and AlGaN growth or at 250W and 2.0 SCCM for InN growth. Films were monitored during growth with a reflection high energy electron diffraction (RHEED) system operated at 20 kV beam voltage.

Homoepitaxial AlN buffer layers, 100 nm thick, were grown on the AlN templates to provide a smooth, low-defect surface for AlGaN film growth. The AlN and AlGaN films were grown under metal-rich regimes at substrate temperatures of 800-900°C. The compositions of the AlGaN films were controlled by varying the Al flux while maintaining an excess of Ga for (Ga+Al)/N = 1.5-2. After growth, the excess Ga droplets were desorbed by increasing the substrate temperature by 50°C and maintaining this temperature until the RHEED intensity reached a plateau.

A series of InN cap layers were grown on AlGaN films across a range of compositions from 30-70% Al. The substrate temperature optimization for InN growth was informed by previous studies which show the ideal temperature window for high adatom mobility without competition with InN decomposition (400-470°C).[26,39–41] InN films were also grown at much lower temperatures, as low as 0°C, to encourage the formation of amorphous material, however all films were crystalline. Therefore, we chose conditions to optimize the crystal quality and minimize extended defects that could act as diffusion pathways. The surface was monitored via RHEED during growth to ensure the film was relatively smooth without significant 3D island formation or indium droplet accumulation. A representative RHEED pattern for a grown InN film is shown in Figure 1. The surface coverage and crystalline quality of the InN did not vary significantly with the composition or thickness of the underlying AlGaN films. All relaxation occurred within the first few monolayers regardless of Al composition. Therefore, the compositional variation in the capped AlGaN films will not be discussed further.

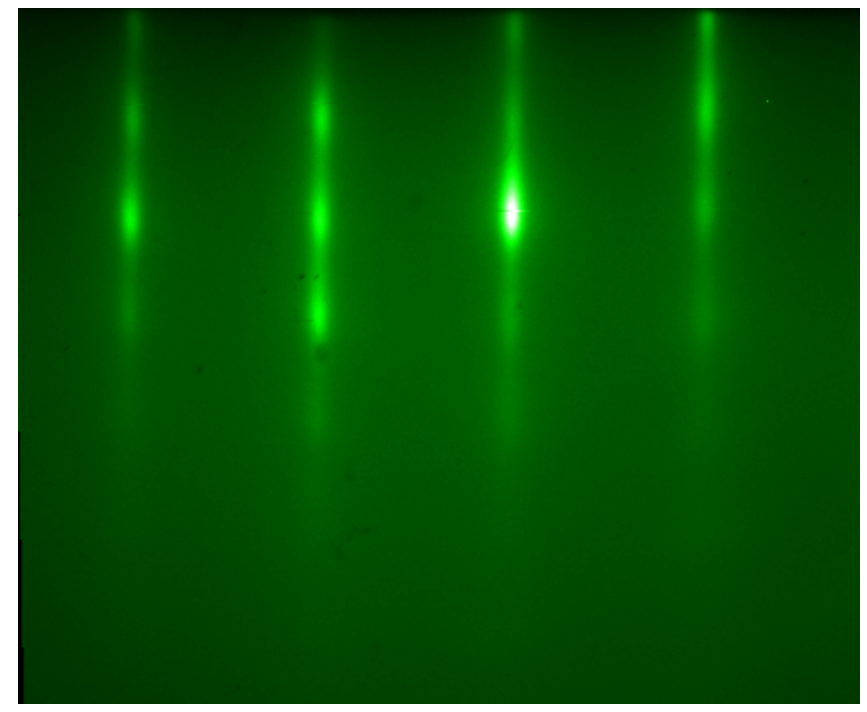

**Figure 1.** Representative RHEED pattern for an InN film grown on AlGaN.

Grown films were characterized by collecting symmetric X-ray diffraction (XRD) patterns using a Rigaku Smartlab diffractometer equipped with Cu Kα radiation and a Ge 2-bounce monochromator. The XRD for a representative sample shown in Figure 2 confirms 0001-oriented wurtzite InN grown on the AlGaN film without the formation of metallic In droplets ($2\theta = 33.0°$). Laue oscillations extending from the InN 0002 reflection suggest high crystalline quality and were used to measure film thicknesses, giving an InN growth rate of ~1.11 Å/s.

Atomic force microscopy (AFM) measurements were collected on the grown films using a Bruker Dimension Icon operated in tapping mode to probe surface morphology. High resolution scanning electron microscope (SEM) images of the InN film surface were collected on a Thermo Fisher Scientific Helios 5 PFIB CXe. Energy Dispersive X-ray spectroscopy (EDS) maps and additional SEM imaging of the InN film surface was conducted on a TESCAN Solaris FIB-SEM equipped with an EDAX Octane Elite Super EDS. Lamella preparation for cross-sectional high-angle annular dark-field scanning transmission electron microscopy (HAADF-STEM) imaging was carried out on a Thermo Fisher Scientific Helios 5 Laser Hydra Plasma FIB-SEM using argon ions in an inverted lift out technique chosen to eliminate Ga alloying artifacts and minimize mechanical warping in very thin specimens.[42] The prepared lamella was imaged using a Thermo Fisher Scientific Spectra 200 STEM operated at a 200 kV accelerating voltage with a convergence semi-angle of 24.2 mrad. STEM-EDS maps were acquired using a four segment Thermo Fisher Scientific SuperX EDS detector.

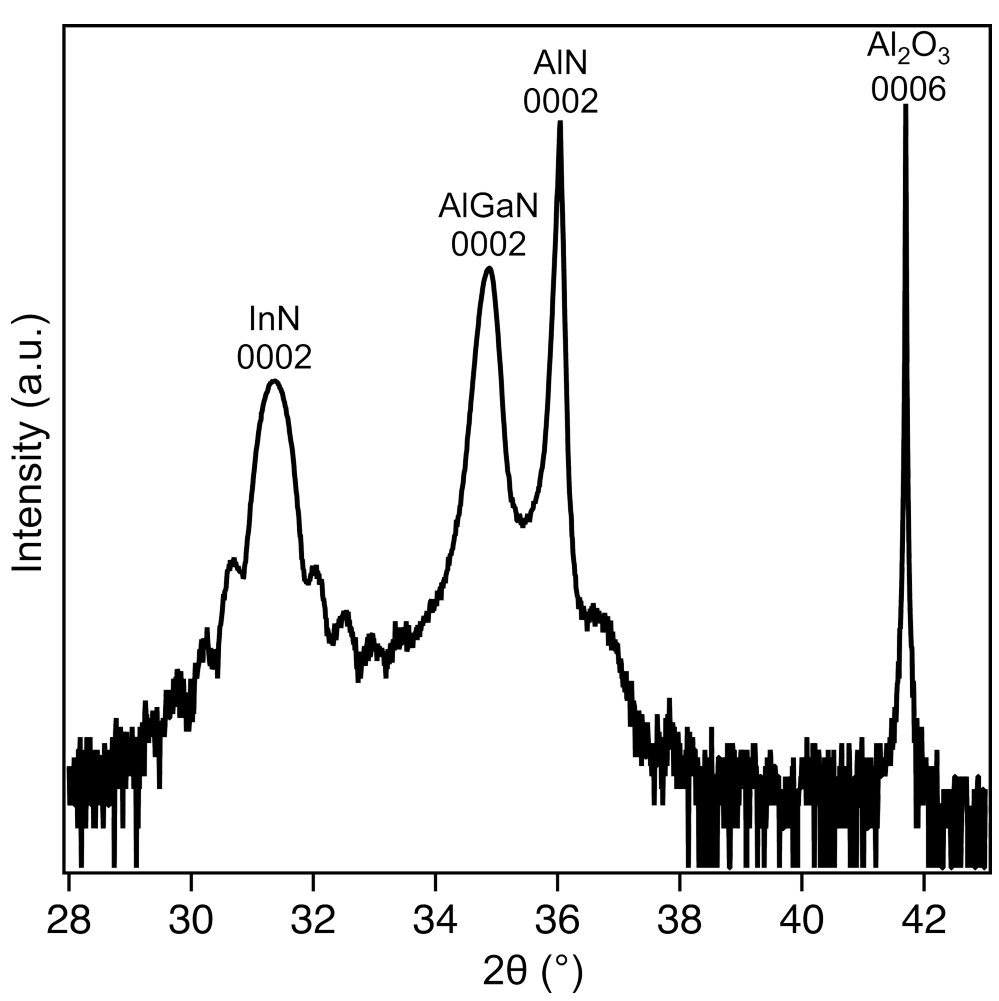

**Figure 2.** XRD pattern showing the 000ℓ peaks for the InN cap layer, the grown AlGaN layer, the AlN template, and the $Al_2O_3$ substrate.

## Decapping and surface studies

Decapping experiments were carried out first in the MBE chamber after depositing an InN cap layer on AlGaN then allowing the chamber to pump out overnight to pressures in the $10^{-10}$ Torr range. Desorbed gases were monitored by a Stanford Research Systems residual gas analyzer (RGA) on a remote port with a background nitrogen signal of ~$1x10^{-11}$ Torr after the overnight chamber pump. The substrate was heated from 330-680°C at a ramp rate of 6°C/min while monitoring the nitrogen signal with the RGA and tracking the RHEED intensity. The surface composition after decapping was examined with x-ray photoelectron spectroscopy (XPS) measurements collected on a Physical Electronics Phi VersaProbe III instrument equipped with a monochromatic Al-kα source (hn = 1486.7 eV).

The decapping experiment was repeated after removing the capped AlGaN films from vacuum and leaving them exposed to atmosphere for time periods ranging from several hours to a few days. The films were placed back under vacuum for collection of XPS measurements during *in situ* heating at a ramp rate of 5°C/min from room temperature up to 750°C. The temperature of the heated stage on the XPS was measured via an integrated Type K thermocouple junction. During *in situ* heating, XPS spectra were acquired at a pass energy of 140 eV.

Additional photon energy-dependent XPS measurements were performed at 29-ID of the Advanced Photon Source at Argonne National Laboratory. Substrates were mounted to a Mo sample plate with Ta clips for *in situ* heating under vacuum. The temperature of the stage was measured with a thermocouple mounted near the sample plate that was calibrated with an optical pyrometer to monitor the temperature emissivity (0.3). Samples were decapped by ramping from room temperature to 570°C over ~20 min and then holding at 570°C for 40 minutes. The composition of the sample surface was measured with XPS after thermal removal of the cap layer.

## 3. RESULTS

### Homogeneity and structure of InN Cap

The surface of the grown InN cap layer was examined by SEM imaging as shown in Figure 3. The images show a mostly homogeneous surface, with small (<0.5 μm) droplets dispersed across the film area separated by approximately 10 – 50 μm. Terraces are visible underneath the grains shown in Fig. 3a, which are residual from step-flow growth in the AlGaN layer and corroborate the spotty plus streaky RHEED images of the InN layer shown in Figure 1. EDS measurements (Figure S1) of the droplets showed that they were either InN islands or metallic In. Surface roughness was characterized by AFM, which gave a roughness of 2.8 nm RMS over a 10 μm$^2$ area (Figure 3c).

The deposited InN cap provided coverage over the full area of the AlGaN film, as demonstrated by the cross-section HAADF-STEM image and EDS elemental maps shown in Figure 4. The cross-section image shows the thin capping layer is uniform across the AlGaN surface and that there is no intermixing into the AlGaN. Carbon is only present on the surface, within the electron beam deposited Pt protective layer. Although there are pits and islands visible at the InN surface, the <10 nm pits do not extend far enough into the ~30 nm thick InN film to expose any of the AlGaN film. Therefore, we suggest that >20 nm thick InN cap layers are effective in shielding underlying material from atmospheric exposure.

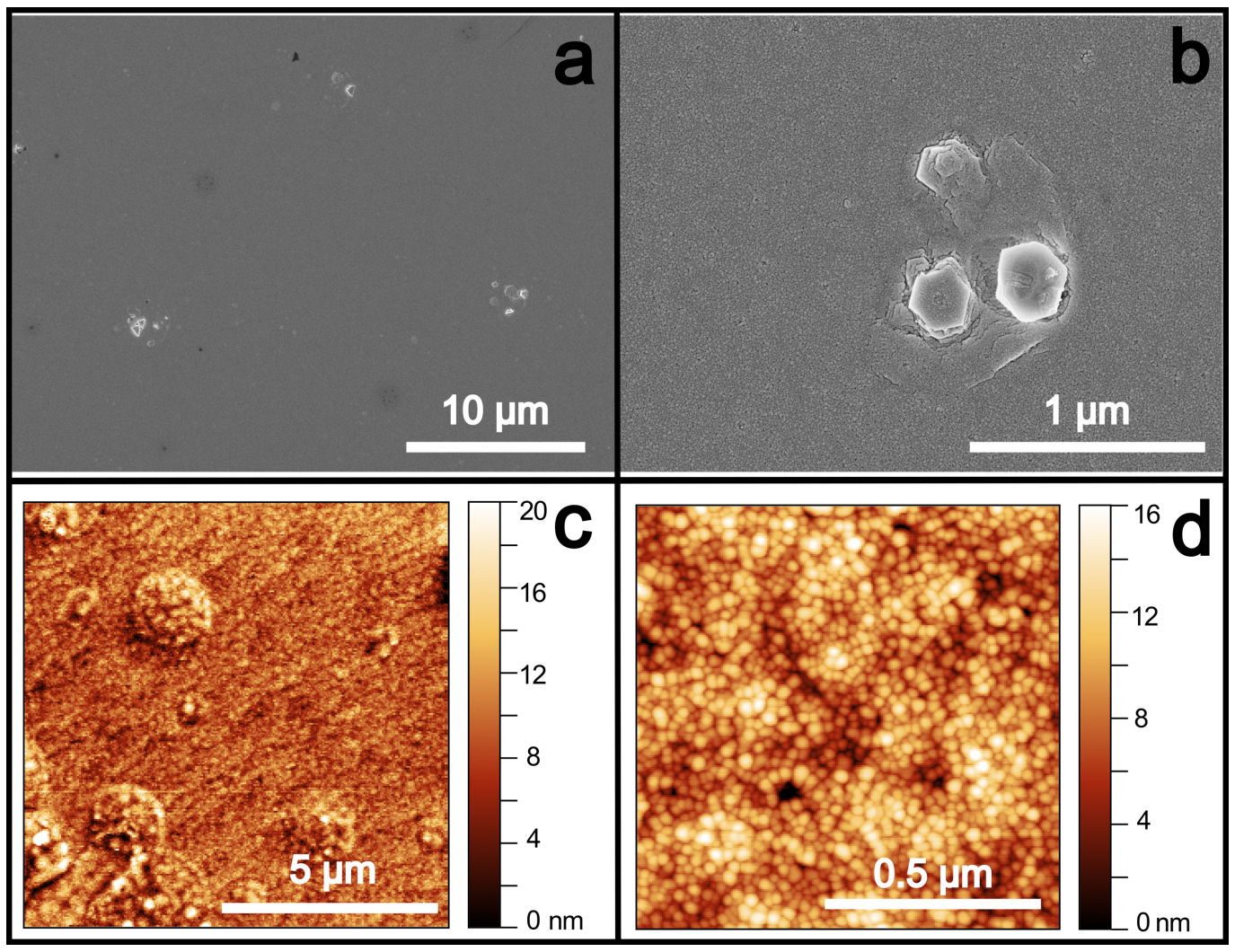

**Figure 3.** Top-down Secondary Electron SEM images collected at 5 kV, 25 pA showing a) a larger area across the film surface and b) magnified view of defects/crystallites dispersed across the surface. c) Representative AFM images of the InN surface at 10 μm$^2$ and 1 μm$^2$ length scales.

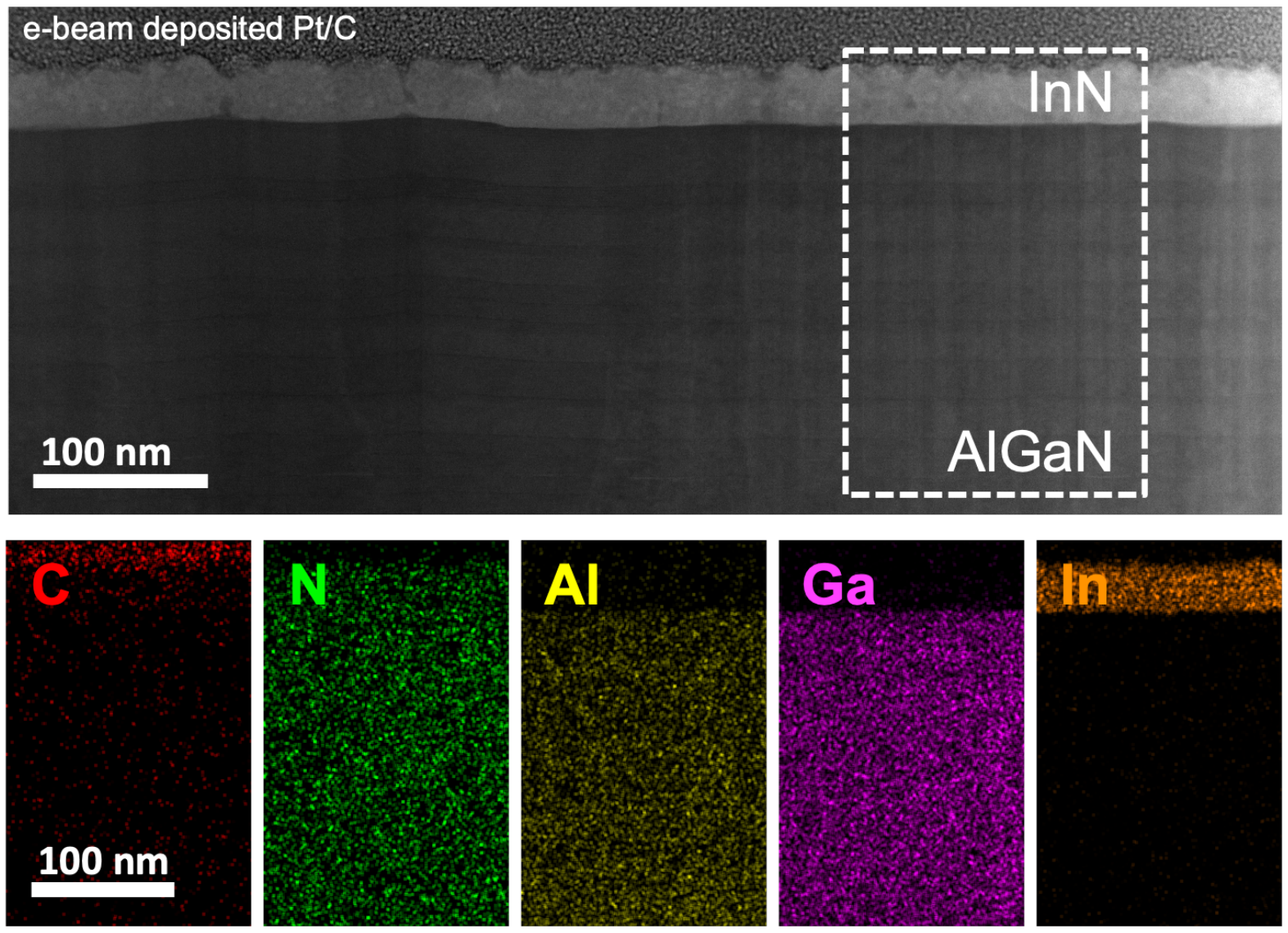

**Figure 4.** Cross-section HAADF-STEM image and EDS elemental maps showing the AlGaN film capped with a thin InN layer. Above the InN layer is a Pt/C layer deposited by e-beam to protect from surface damage while thinning the lamella.

## Thermal removal

The InN cap was desorbed *in situ* in the MBE chamber by heating the substrate. The decomposition and desorption of the InN cap was characterized by simultaneously monitoring the $N_2^-$ signal (mass-to-charge ratio 28) via RGA and tracking the RHEED intensities of the InN then AlGaN films, with results shown in Figure 5. The InN removal takes place by first decomposing InN into In and $N_2$, as indicated by the increase in nitrogen signal and decrease in RHEED intensity at 550°C.[26] After full decomposition, the RHEED intensity reaches a minimum due to the presence of adsorbed metallic In on the surface. The desorption of metallic In begins around 615°C, and the RHEED intensity for the exposed AlGaN film increases until desorption is complete at approximately 680°C. The RGA nitrogen signal remains elevated after decomposition due to heater outgassing from the temperature ramp profile.

After decapping, the AlGaN sample was transferred through atmosphere before XPS measurements of the composition of the film surface to quantify desorption of the InN, with the surface composition of the decapped film shown in Table S1. The measured In concentration is below the noise floor, indicating that the InN cap is fully removable. Additionally, the high percentage of carbon and oxygen atoms present at the AlGaN surface after brief atmospheric exposure highlights the importance of protective capping to prevent contamination. The decap procedure was replicated for a capped film that had been left in air for several days, indicating that exposure to atmosphere does not inhibit the desorption of In or increase the concentration of non-desorbing species.

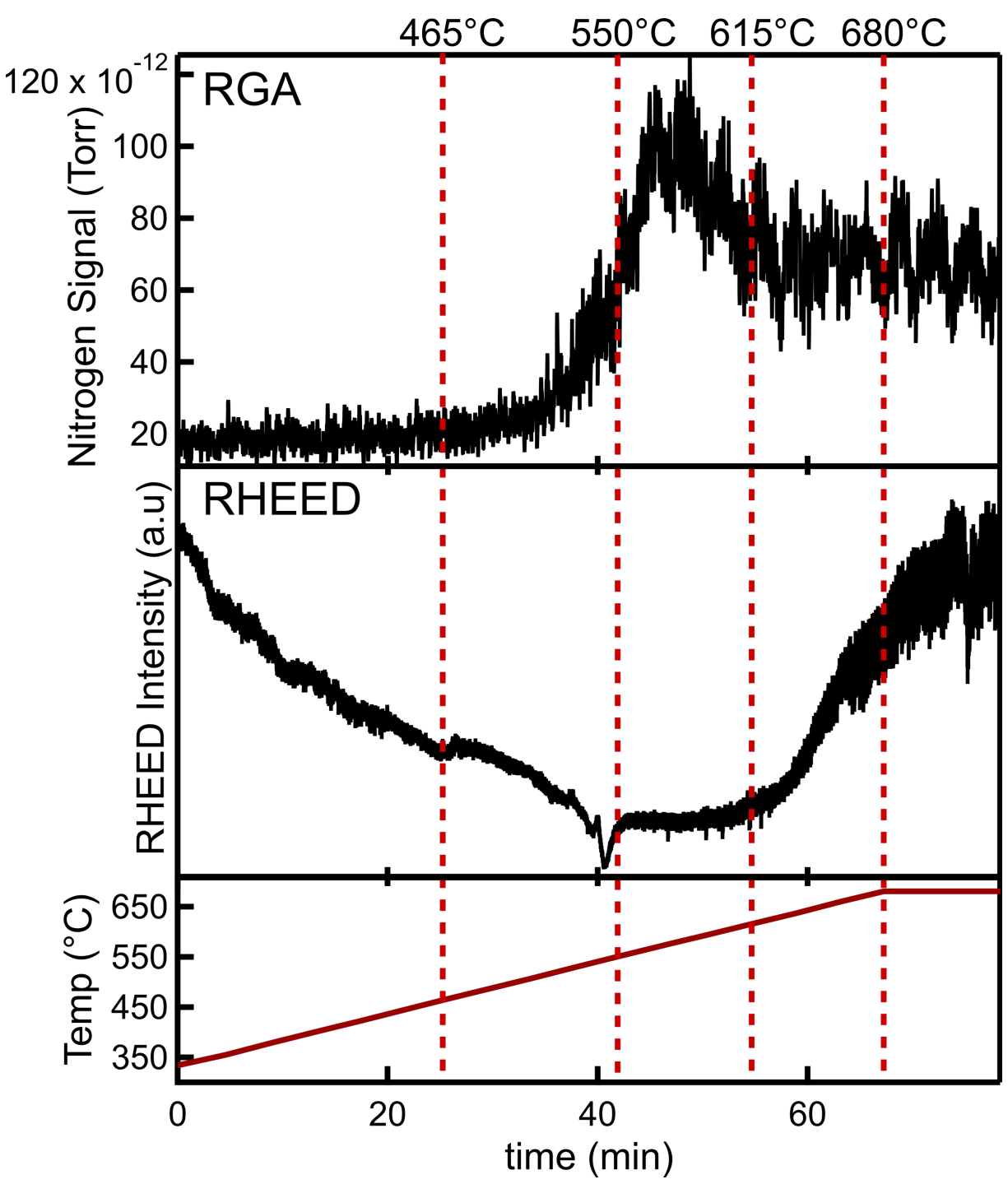

**Figure 5.** RHEED intensity and nitrogen signal from RGA as a function of time as temperature is ramped at 6°C/min from 330-680°C.

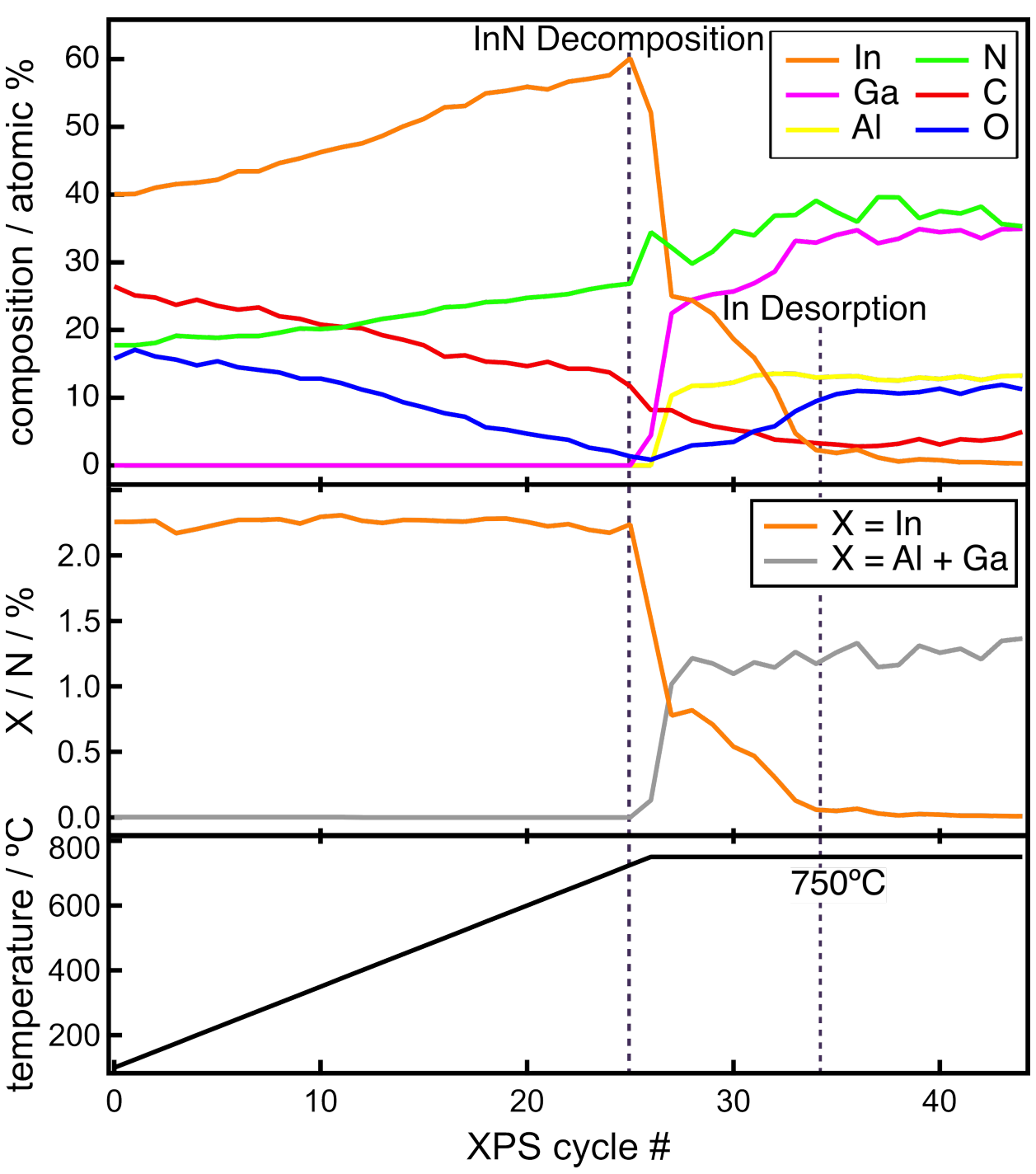

**Figure 6.** Composition determined from XPS as a function of heating at a ramp rate of 5°C/min up to 750°C for InN cap removal.

**Table 1.** Film surface compositions measured by XPS after decapping of InN layer in the XPS system

| Element | C 1s | O 1s | N 1s | Ga 3d | Al 2p | In 3d 5/2 |
|---|---|---|---|---|---|---|
| Approx. atomic % | 3.9 | 11.3 | 37.3 | 34.2 | 12.9 | 0.4 |

Removal of the InN cap layer was then carried out *in situ* in the XPS chamber to track the surface composition as a function of annealing temperature, with results shown in Figure 6. The decomposition of the InN layer is characterized by the rapid decrease in In atomic % and increase in Al and Ga atomic % at temperatures near 750°C. The desorption of the remaining indium occurs while holding the temperature at 750°C. The composition at the surface after the In is fully desorbed is summarized in Table 1. Importantly, the carbon and oxygen percentages reach minima near the noise floor during the decapping process, with C at a minimum after In desorption and O at a minimum after InN decomposition. However, the oxygen signal increases during In desorption because the exposed AlGaN surface begins to react with residual oxygen-containing species that are likely degassing from the XPS chamber while heating, which is evidenced by the rise in pressure from 1E-9 Torr to 1E-8 Torr during the first 2 minutes (approximately 6 XPS cycles) of the 750°C temperature dwell.

The reaction of AlGaN with residual oxygen-containing species was investigated by sputter cleaning a decapped AlGaN film to remove all surface oxygen before heating. Oxidation was again observed at high temperatures, indicating an external contamination source (SI Figure S2). Photon energy-dependent XPS measurements performed at 29-ID of Advanced Photon Source at Argonne National Laboratory confirm that the surface oxygen percentage is dependent on the partial pressure of oxygen in the vacuum system. This indicates that contamination from *atmospheric exposure* is contained within the InN cap layer and emphasizes the importance of reducing potential sources of residual oxygen-containing species by using cap layers that are not oxyphilic.

## 4. DISCUSSION

While the InN capping layer contains small islands of InN or In droplets dispersed across the surface, the coverage across the film area is consistent enough to fully protect the film beneath from exposure to atmosphere. The desorption experiments reveal the two-step process by which InN is fully removed by heating, even after prolonged exposure to atmosphere. The InN film first undergoes decomposition into In and $N_2$ before remaining metallic In atoms at the film surface are desorbed at higher temperatures. The XPS desorption experiments showed that carbon and oxygen contamination from atmospheric exposure was present in the InN layer but did not extend to the AlGaN surface, with the contaminants reaching their minima when the InN cap was removed.

Because the decapping temperature for InN is far below the decomposition temperature for other III-nitrides including AlGaN, its removal was not expected to alter the AlGaN surface structure. This was confirmed by comparing the RHEED patterns of the AlGaN film before and after InN deposition and removal, as shown in Figure 7. The surface of both unprotected AlGaN films and AlGaN films after InN cap removal were examined with SEM imaging (SI Figure S4). The unprotected AlGaN film surface displayed a higher density of pits and/or islands dispersed across a given area vs the AlGaN film that had been decapped. The lower density of large defects across the film area and a more uniform contrast in the SEM images suggest that the capping and decapping process does not damage the AlGaN surface and may actually promote a smoother surface.

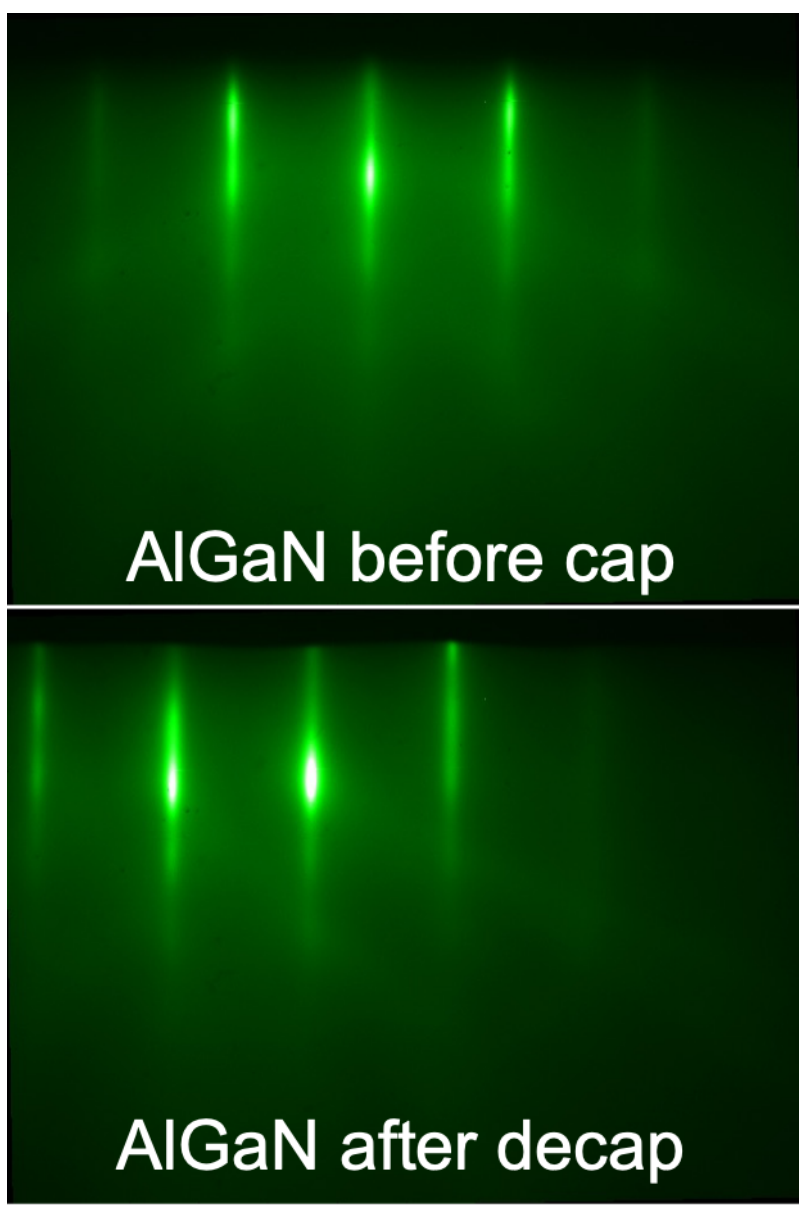

**Figure 7.** RHEED patterns for deposited AlGaN film before InN cap deposition and after removal of the InN cap by heating.

### Isoenergy surface of decapped AlGaN

Capped samples were transported to the Advanced Photon Source at Argonne National Laboratory for *in situ* decapping. The removal of the InN cap was confirmed by XPS measurements of the sample surface composition, as shown in SI Figure S3. After thermal removal of the InN cap, the film surface contained less than 5% indium, and the oxygen content was dependent on the background partial pressure of oxygen in the vacuum system. ARPES measurements only resulted in the observation of band dispersions when there was minimal oxygen present. Soft X-ray ARPES measurements are highly sensitive to even sub-monolayer adsorbates and are therefore a highly sensitive probe of the success of the cap/decap procedure for contamination mitigation.[25]

The AlGaN surface was fully recovered following decapping. This was confirmed by mapping the band dispersion using ARPES. The isoenergy surface measurement taken at 750 eV with circular polarization reveals the high symmetry directions of $Al_{0.4}Ga_{0.6}N$, with the Brillouin zone is shown in red for reference, in plane lattice vector of 3.16 Å, the weighted average of AlN and GaN lattice parameters. The valence band dispersion was collected at both room temperature and 93K along the direction indicated by the dashed line through the zone corners, as shown in Figure 8. Undoped ultra-wide bandgap semiconductors with low conductivity, such as the AlGaN demonstrated here, are extremely sensitive to surface contamination and disorder at the surface. The observation of bands, even with weak signal, supports the effectiveness of this capping layer, and the weak signal is primarily due to cation disorder in the $Al_{0.4}Ga_{0.6}N$ random alloy. The expected inverted parabolic energy-momentum dispersion around Γ was also resolved, which qualitatively matches density functional theory calculations.[43] The effectiveness of the InN cap for surface protection was further evidenced by attempts to take ARPES measurements of similar AlGaN films that had not been capped, where band dispersion could not be observed due to signal saturation by surface contaminants.

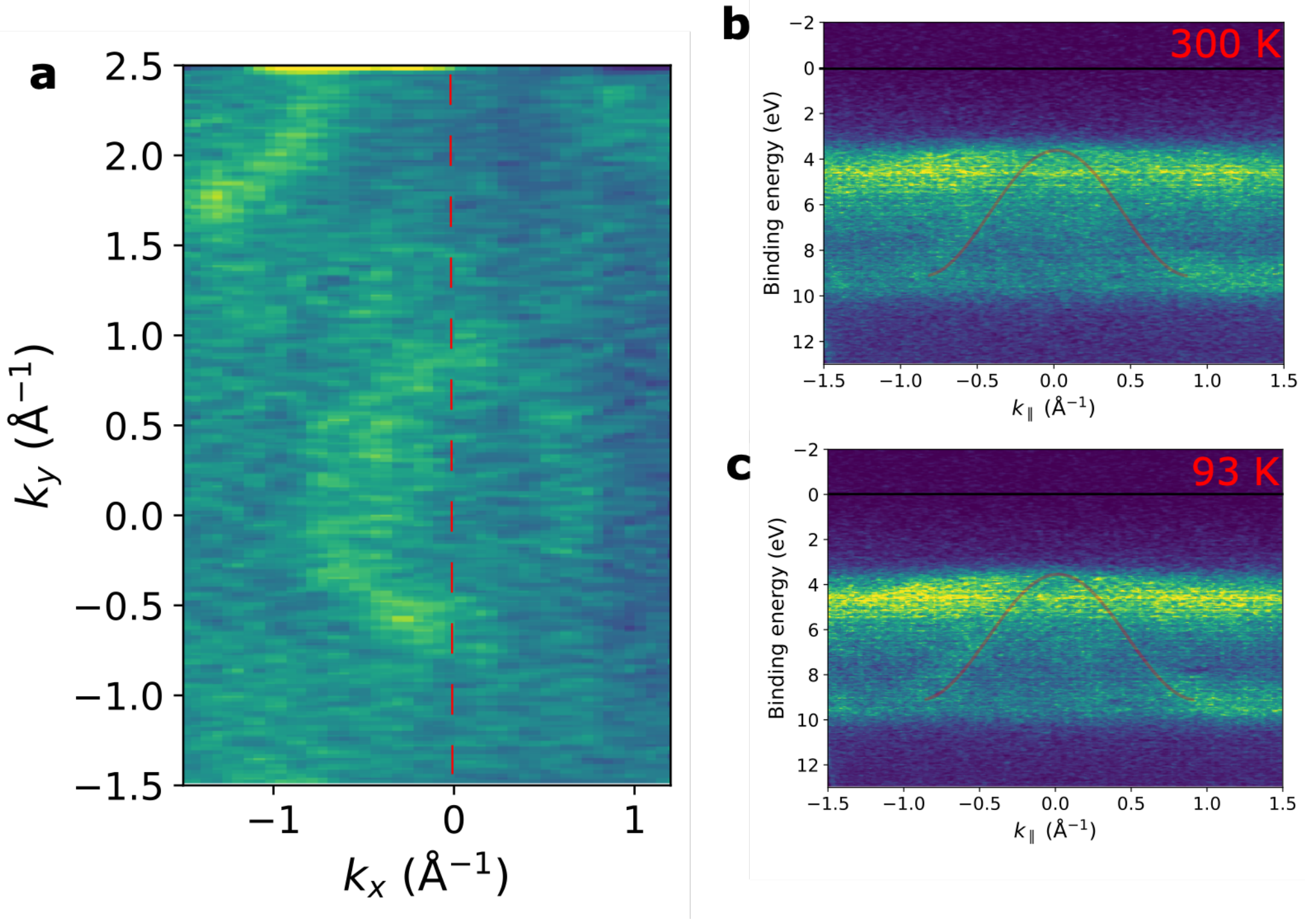

**Figure 8.** a) Room temperature isoenergy surface (4.7 eV below the valence band maxima of the decapped sample taken at photon energy of 750 eV). b,c) Band dispersions along the high symmetry direction (dashed line shown in (a)) spanning the zone corners at 300K temperature and at 93 K respectively. The transparent red curve traces the inverted parabolic hole-like dispersion around Γ.

## 5. CONCLUSIONS

We have demonstrated a reversible InN capping layer for III-N and related materials to protect films of interest from contamination during transfer through air for processing and characterization. This opens up opportunities for characterization via surface-sensitive techniques that are otherwise unachievable due to the overwhelming effect of contaminants, as demonstrated by ARPES measurements after *in situ* decapping of the AlGaN films resulting in the observation of band dispersions. While AlGaN was chosen as a model system, this procedure is generalizable and can be replicated with a variety of different materials that have higher decomposition temperatures than InN including ferroelectric AlN alloys,[32–36] transition metal nitrides,[37,38] and carbides of interest as conductive substrates for AlGaN, such as TaC.[44] Development of InN as a protective cap layer on a variety of films will enable future surface and interface studies with surface-sensitive techniques, aiding in the advancement of an understanding of the fundamental physics of surface effects.

## ASSOCIATED CONTENT

The supporting information is available free of charge.

The SI includes SEM-EDS maps of In or InN droplets on film surfaces, the XPS composition measured after decapping a film in the MBE and transporting through air, XPS data showing the oxidation of an AlGaN surface during heating, an XPS spectra taken at the Advanced Photon Source after decapping, and SEM images of the surfaces of AlGaN films that were either not capped with InN or capped and decapped.

## AUTHOR INFORMATION

### Corresponding Author

M. Brooks Tellekamp, brooks.tellekamp@nlr.gov

### Author Contributions

The manuscript was written through contributions of all authors. All authors have given approval to the final version of the manuscript.

### Funding Sources

This research was funded by the U.S. Department of Energy (DOE), Office of Science, Basic Energy Sciences.

## ACKNOWLEDGMENT

The work was solely supported as part of A Center for Power Electronics Materials and Manufacturing Exploration (APEX), an Energy Frontier Research Center funded by the U.S. Department of Energy, Office of Science, Basic Energy Sciences. This work was authored in part by the National Laboratory of the Rockies (NLR) for the U.S. Department of Energy (DOE), operated under Contract No. DE-AC36-08GO28308. This research was performed in part on APS beam time award(s) (DOI: https://doi.org/10.46936/APS-193816/60016732) from the Advanced Photon Source, a U.S. Department of Energy Office of Science user facility at Argonne National Laboratory. The views expressed in the article do not necessarily represent the views of the DOE or the U.S. Government.

## ABBREVIATIONS

ARPES, Angle-resolved photoemission spectroscopy; 2DEG, two-dimensional electron gas; PA-MBE, plasma-assisted molecular beam epitaxy; RHEED, reflection high energy electron diffraction; EDS, energy dispersive x-ray spectroscopy; AFM, atomic force microscopy; XRD, x-ray diffraction; HAADF-STEM, high-angle annular dark-field scanning transmission electron microscopy; XPS, x-ray photoelectron spectroscopy; RGA, residual gas analyzer

## REFERENCES

(1) Collaert, N.; Alian, A.; Banerjee, A.; Boccardi, G.; Cardinael, P.; Chauhan, V.; Desset, C.; ElKashlan, R.; Khaled, A.; Kunert, B.; Mols, Y.; O'Sullivan, B.; Peralagu, U.; Pinho, N.; Rodriguez, R.; Sibaja-Hernandez, A.; Sinha, S.; Sun, X.; Vais, A.; Vermeersch, B.; Yadav, S.; Yan, D.; Yu, H.; Zhang, Y.; Zhao, M.; Driessche, J. V.; Gramegna, G.; Wambacq, P.; Parvais, B.; Peeters, M. III-V/III-N Technologies for next Generation High-Capacity Wireless Communication. *2022 Int. Electron Devices Meet. IEDM* **2022**, 11.5.1-11.5.4. https://doi.org/10.1109/IEDM45625.2022.10019555

(2) Baten, Z.; Alam, S.; Sikder, B.; Aziz, A. III-Nitride Light-Emitting Devices. *Photonics* **2021**, *8* (430). https://doi.org/10.3390/photonics8100430

(3) Agrawal, S.; Van Deurzen, L.; Encomendero, J.; Dill, J. E.; Wei (Sheena) Huang, H.; Protasenko, V.; Xing, H. (Grace); Jena, D. Ultrawide Bandgap Semiconductor Heterojunction p–n Diodes with Distributed Polarization-Doped p-Type AlGaN Layers on Bulk AlN Substrates. *Appl. Phys. Lett.* **2024**, *124* (10), 102109. https://doi.org/10.1063/5.0189419

(4) Chow, T. P.; Tyagi, R. Wide Bandgap Compound Semiconductors for Superior High-Voltage Unipolar Power Devices. *IEEE Trans. Electron Devices* **1994**, *41* (8),

1481–1483. https://doi.org/10.1109/16.297751
(5) Khan, M. A.; Kuznia, J. N.; Van Hove, J. M.; Pan, N.; Carter, J. Observation of a Two-Dimensional Electron Gas in Low Pressure Metalorganic Chemical Vapor Deposited GaN-Al *x* Ga1− *x* N Heterojunctions. *Appl. Phys. Lett.* **1992**, *60* (24), 3027–3029. https://doi.org/10.1063/1.106798
(6) Kocan, M.; Rizzi, A.; Luth, H.; Keller, S.; Mishra, U. K. Surface Potential at As-Grown GaN(0001) MBE Layers. *Phys. Status Solidi B* **2002**, *234* (3), 773–777. https://doi.org/10.1002/1521-3951(200212)234:3%3C773::AID-PSSB773%3E3.0.CO;2-0
(7) Rizzi, A.; Kocan, M.; Malindretos, J.; Schildknecht, A.; Teofilov, N.; Thonke, K.; Sauer, R. Surface and Interface Electronic Properties of AlGaN(0001) Epitaxial Layers. *Appl. Phys. A* **2007**, *87* (3), 505–509. https://doi.org/10.1007/s00339-007-3873-4
(8) Benjamin, M. C.; Bremser, M. D.; Weeks, T. W.; King, S. W.; Davis, R. F.; Nemanich, R. J. UV Photoemission Study of Heteroepitaxial A1GaN Films Grown on 6H-SiC. *Appl. Surf. Sci.* **1996**, *104/105*, 455–460.
(9) Qin, X.; Lucero, A.; Azcatl, A.; Kim, J.; Wallace, R. M. *In Situ* x-Ray Photoelectron Spectroscopy and Capacitance Voltage Characterization of Plasma Treatments for Al2O3/AlGaN/GaN Stacks. *Appl. Phys. Lett.* **2014**, *105* (1), 011602. https://doi.org/10.1063/1.4887056
(10) Lev, L. L.; Maiboroda, I. O.; Husanu, M.-A.; Grichuk, E. S.; Chumakov, N. K.; Ezubchenko, I. S.; Chernykh, I. A.; Wang, X.; Tobler, B.; Schmitt, T.; Zanaveskin, M. L.; Valeyev, V. G.; Strocov, V. N. K-Space Imaging of Anisotropic 2D Electron Gas in GaN/GaAlN High-Electron-Mobility Transistor Heterostructures. *Nat. Commun.* **2018**, *9* (1), 2653. https://doi.org/10.1038/s41467-018-04354-x
(11) Lippert, G.; Krüger, D.; Zeindl, H. P.; Ramm, J.; Bugiel, E.; Osten, H. J. Problems Of Contamination Prior And During SI-MBE. *MRS Online Proc. Libr.* **1993**, *315* (1), 85–90. https://doi.org/10.1557/PROC-315-85
(12) Hey, R.; Wassermeier, M.; Höricke, M.; Wiebicke, E.; Kostial, H. Minimizing Interface Contamination in MBE Overgrowth. *J. Cryst. Growth* **1999**, *201–202*, 582–585. https://doi.org/10.1016/S0022-0248(98)01413-4
(13) Arisio, C.; Cassou, C. A.; Lieberman, M. Loss of Siloxane Monolayers from GaN Surfaces in Water. *Langmuir* **2013**, *29*, 5145–5149. https://doi.org/10.1021/la400849j
(14) Greczynski, G. Impact of Sample Storage Type on Adventitious Carbon and Native Oxide Growth: X-Ray Photoelectron Spectroscopy Study. **2022**, *205*, 111463. https://doi.org/10.1016/j.vacuum.2022.111463
(15) Parish, G.; Keller, S.; Denbaars, S. P.; Mishra, U. K. SIMS Investigations into the Effect of Growth Conditions on Residual Impurity and Silicon Incorporation in GaN and AlxGa1–xN. *J. Electron. Mater.* **2000**, *29* (1).
(16) McCandless, J. P.; Gorsak, C. A.; Protasenko, V.; Schlom, D. G.; Thompson, M. O.; Xing, H. G.; Jena, D.; Nair, H. P. Accumulation and Removal of Si Impurities on B-Ga2O3 Arising from Ambient Air Exposure. *Appl. Phys. Lett.* **2024**, *124*, 111601.
(17) Koblmüller, G.; Chu, R. M.; Raman, A.; Mishra, U. K.; Speck, J. S. High-Temperature Molecular Beam Epitaxial Growth of AlGaN/GaN on GaN Templates with Reduced Interface Impurity Levels. *J Appl Phys* **2010**, *107*, 043527. https://doi.org/10.1063/1.3285309
(18) King, S. W.; Barnak, J. P.; Bremser, M. D.; Tracy, K. M.; Ronning, C.; Davis, R. F.; Nemanich, R. J. Cleaning of AlN and GaN Surfaces. *J. Appl. Phys.* **1998**, *84* (9), 5248–5260. https://doi.org/10.1063/1.368814
(19) Inagaki, T.; Hashizume, T.; Hasegawa, H. Effects of Surface Processing on 2DEG Current Transport at AlGaN/GaN Interface Studied by Gateless HFET Structure. *Appl. Surf. Sci.* **2003**, *216* (1–4), 519–525.

https://doi.org/10.1016/S0169-4332(03)00482-3
(20) Taglauer, E. Surface Cleaning Using Sputtering. *Appl. Phys. Solids Surf.* **1990**, *51* (3), 238–251. https://doi.org/10.1007/BF00324008
(21) Kowalczyk, S. P.; Miller, D. L.; Waldrop, J. R.; Newman, P. G.; Grant, R. W. Protection of Molecular Beam Epitaxy Grown Al *x* Ga1− *x* As Epilayers during Ambient Transfer. *J. Vac. Sci. Technol.* **1981**, *19* (2), 255–256. https://doi.org/10.1116/1.571114
(22) Karpov, I.; Venkateswaran, N.; Bratina, G.; Gladfelter, W.; Franciosi, A.; Sorba, L. Arsenic Cap Layer Desorption and the Formation of GaAs(001)c(4x4) Surfaces. *J Vac Sci Technol B* **2026**, *13* (5).
(23) Resch, U.; Esser, N.; Raptis, Y. S.; Richter, W.; Wasserfall, J.; Förster, A.; Westwood, D. I. Arsenic Passivation of MBE Grown GaAs(100): Structural and Electronic Properties of the Decapped Surfaces. *Surf. Sci.* **1992**, *269–270* (15), 797–803. https://doi.org/10.1016/0039-6028(92)91351-B
(24) Resch, U.; Scholz, S. M.; Rossow, U.; Mtiller, A. B.; Richter, W. Thermal Desorption of Amorphous Arsenic Caps from GaAs(100) Monitored by Reflection Anisotropy Spectroscopy. *Appl. Surf. Sci.* **1993**, *63*, 106–110.
(25) Hajlaoui, M.; Ponzoni, S.; Deppe, M.; Henksmeier, T.; As, D. J.; Reuter, D.; Zentgraf, T.; Springholz, G.; Schneider, C. M.; Cramm, S.; Cinchetti, M. Extremely Low-Energy ARPES of Quantum Well States in Cubic-GaN/AlN and GaAs/AlGaAs Heterostructures. *Sci. Rep.* **2021**, *11* (1), 19081. https://doi.org/10.1038/s41598-021-98569-6
(26) Gallinat, C. S.; Koblmüller, G.; Brown, J. S.; Speck, J. S. A Growth Diagram for Plasma-Assisted Molecular Beam Epitaxy of In-Face InN. *J. Appl. Phys.* **2007**, *102* (6), 064907. https://doi.org/10.1063/1.2781319
(27) Pisch, A.; Schmid-Fetzer, R. In Situ Decomposition Study of GaN Thin Films. *J. Cryst. Growth* **1998**, *187* (3–4), 329–332. https://doi.org/10.1016/S0022-0248(98)00006-2
(28) Choi, H. W.; Rana, M. A.; Chua, S. J.; Osipowicz, T.; Pan, J. S. Surface Analysis of GaN Decomposition. *Semicond. Sci. Technol.* **2002**, *17* (12), 1223–1225. https://doi.org/10.1088/0268-1242/17/12/304
(29) Hildenbrand, D. L.; Hall, W. F. THE VAPORIZATION BEHAVIOR OF BORON NITRIDE AND ALUMINUM NITRIDE[1]. *J. Phys. Chem.* **1963**, *67* (4), 888–893. https://doi.org/10.1021/j100798a041
(30) Sarantopoulou, E.; Kollia, Z.; Dra, G. Long-Term Oxidization and Phase Transition of InN Nanotextures. *Nanoscale Res. Lett.* **2011**, *6*, 387.
(31) Foley, C. P.; Lyngdal, J. Analysis of Indium Nitride Surface Oxidation. *J Vac Sci Technol A* **1987**, *5*, 1708–1712. https://doi.org/10.1116/1.574558
(32) Savant, C.; Singh, R.; Nguyen, T.-S.; Casamento, J.; Xing, H. G.; Jena, D. Self-Activated Growth of Cubic AlScN Films from Molecular Nitrogen without Plasma. *J. Appl. Phys.* **2026**, *139* (7), 075307. https://doi.org/10.1063/5.0322931
(33) Savant, C.; Gund, V.; Nomoto, K.; Maeda, T.; Jadhav, S.; Lee, J.; Ramesh, M.; Kim, E.; Nguyen, T.-S.; Chen, Y.-H.; Casamento, J.; Rana, F.; Lal, A.; Xing, H. G.; Jena, D. Ferroelectric AlBN Films by Molecular Beam Epitaxy. *Appl. Phys. Lett.* **2024**, *125* (7), 072902. https://doi.org/10.1063/5.0181217
(34) Shen, R.; Tanim, M. M. H.; Gan, Y.; Liu, J.; Mondal, S.; Li, Y.; Yang, S.; Deotare, P.; Kira, M.; Sun, K.; Mi, Z. Composition-Dependent Ferroelectricity in Epitaxial AlYN. *Appl. Phys. Lett.* **2026**, *129* (4), 042905. https://doi.org/10.1063/5.0340250
(35) Shen, R.; Tanim, M. M. H.; Liu, J.; Yang, S.; Gan, Y.; Li, Y.; Zhang, J.; Nguyen, H.; Zhang, X.; Kioupakis, E.; Deotare, P.; Kira, M.; Sun, K.; Mi, Z. Molecular Beam Epitaxy and Characterization of Wurtzite LaAlN. *Appl. Phys. Lett.* **2026**, *129* (6), 061906. https://doi.org/10.1063/5.0344771
(36) Singh, R.; Bhattacharya, D.; Prakash Savant, C.; Ramesh, M.; Veeraraghavan, N.; Nguyen, T.-S.; Xing, H. G.; Jena, D. Annealing Reduces Leakage Current in MBE Grown AlScN and AlYN Films. *J. Appl. Phys.* **2026**, *140* (4), 045501. https://doi.org/10.1063/5.0331208

(37) Katzer, D. S.; Nepal, N.; Hardy, M. T.; Downey, B. P.; Storm, D. F.; Jin, E. N.; Yan, R.; Khalsa, G.; Wright, J.; Lang, A. C.; Growden, T. A.; Gokhale, V.; Wheeler, V. D.; Kramer, A. R.; Yater, J. E.; Xing, H. G.; Jena, D.; Meyer, D. J. Molecular Beam Epitaxy of Transition Metal Nitrides for Superconducting Device Applications. *Phys. Status Solidi A* **2020**, *217* (3), 1900675. https://doi.org/10.1002/pssa.201900675

(38) Bi, J.; Zhang, R.; Yao, X.; Cao, Y. The Rise of Refractory Transition-Metal Nitride Films for Advanced Electronics and Plasmonics. *Adv. Mater. Interfaces* **2025**, *12* (12), 2500116. https://doi.org/10.1002/admi.202500116

(39) Clinton, E. A.; Vadiee, E.; Tellekamp, M. B.; Doolittle, W. A. Observation and Mitigation of RF-Plasma-Induced Damage to III-Nitrides Grown by Molecular Beam Epitaxy. *J. Appl. Phys.* **2019**, *126* (1), 015705. https://doi.org/10.1063/1.5097557

(40) Jmerik, V. N.; Vekshin, V. A.; Shubina, T. V.; Ratnikov, V. V.; Ivanov, S. V.; Monemar, B. Growth of Optically-active InN with AlInN Buffer by Plasma-assisted Molecular Beam Epitaxy. *Phys. Status Solidi C* **2003**, *0* (7), 2846–2850. https://doi.org/10.1002/pssc.200303520

(41) Feng, Z. C.; Xie, D.; Nafisa, M. T.; Lin, H.-H.; Lu, W.; Chen, J.-M.; Yiin, J.; Chen, K.-H.; Chen, L.-C.; Klein, B.; Ferguson, I. T. Optical, Surface, and Structural Studies of InN Thin Films Grown on Sapphire by Molecular Beam Epitaxy. *J. Vac. Sci. Technol. A* **2023**, *41* (5), 053401. https://doi.org/10.1116/6.0002665

(42) Smeaton, M. A.; Acharya, K.; Sacchi, A.; Gannon, R. N.; Tellekamp, M. B.; Zakutayev, A.; Stevanovic, V.; Spurgeon, S. R. Revealing the Atomic Structure of NiO/Ga2O3 Interfaces. arXiv August 10, 2026. https://doi.org/10.48550/arXiv.2608.10226

(43) Kyrtsos, A.; Matsubara, M.; Bellotti, E. First-Principles Study of the Impact of the Atomic Configuration on the Electronic Properties of Al x Ga 1 − x N Alloys. *Phys. Rev. B* **2019**, *99* (3), 035201. https://doi.org/10.1103/PhysRevB.99.035201

(44) Roberts, D. M.; Hachtel, J. A.; Haegel, N. M.; Miller, M. K.; Rice, A. D.; Tellekamp, M. B. Designing TaC Virtual Substrates for Vertical AlxGa1−xN Power Electronics Devices. *PRX Energy* **2024**, *3*, 033007. https://doi.org/10.1103/PRXEnergy.3.033007

# Supporting Information: Thermally Desorbable InN Capping Layers for Nitride Surface Science

Mellie Lemon,[1] Amitayush Thakur,[2] Anthony Rice,[1] Glenn Teeter,[1] Michelle Smeaton,[1] Renae Gannon,[1] Jessica L. McChesney,[3] M. Brooks Tellekamp[1]

*[1]National Laboratory of the Rockies, Golden, CO, USA*
*[2]Advanced Photon Source, Argonne National Laboratory, Lemont, IL, USA*
*[3]Materials Science Division, Argonne National Laboratory, Lemont, IL, USA*

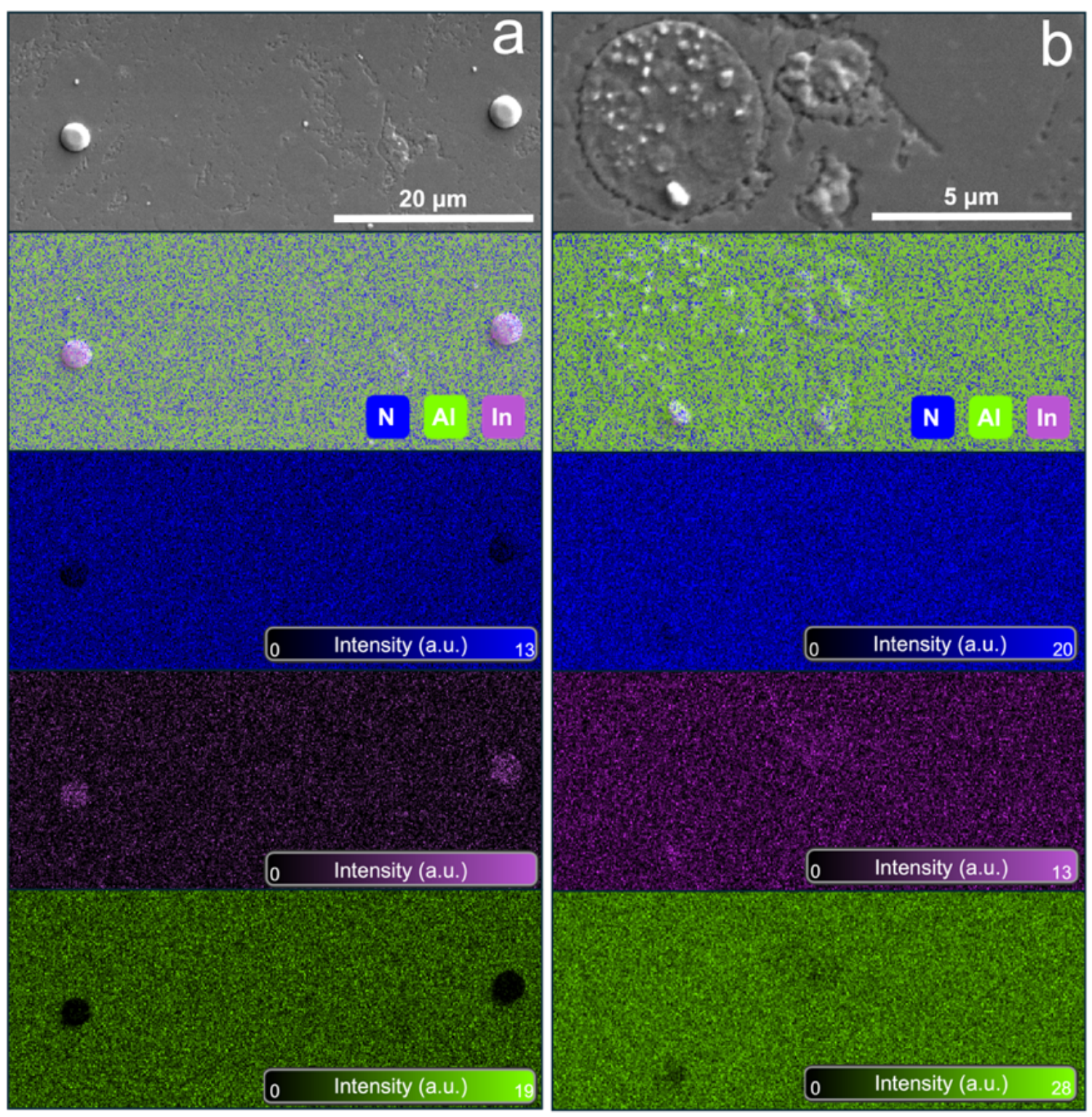

**Figure S1.** EDS maps of InN film surfaces showing either a) In droplets or b) small InN islands. Maps were collected on an EDAX Octane Elite Super EDS detector on a Tescan Solaris FIB-SEM

**Table S1.** Surface composition of film measured by XPS after decapping in MBE and transporting through atmosphere post-decap.

| Element | C 1s | O 1s | N 1s | Ga 3d | Al 2p | In 3d 5/2 |
|---|---|---|---|---|---|---|
| Approx. atomic % | 6.7 | 20.5 | 9.9 | 24.9 | 38.0 | 0.0 |

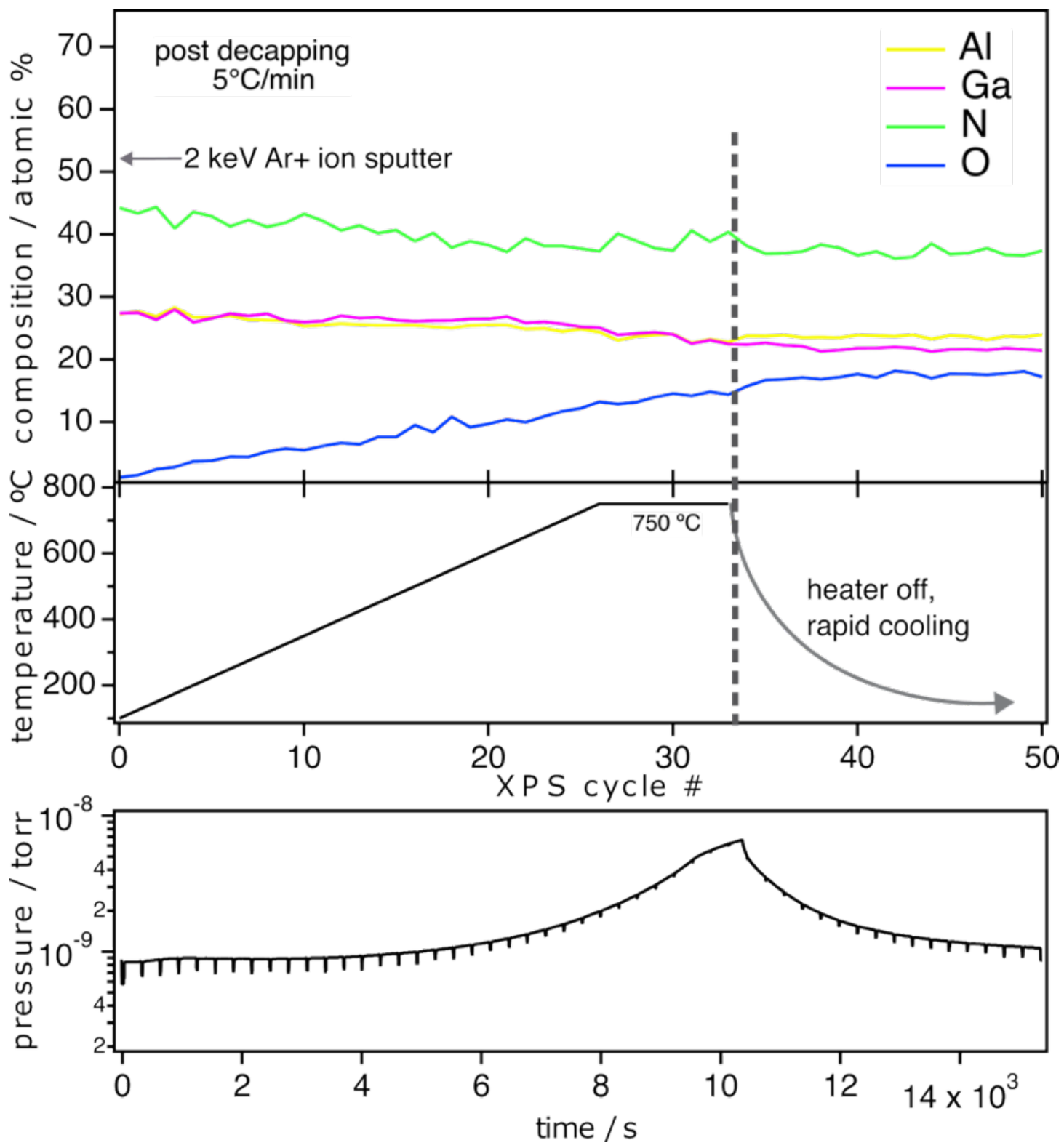

**Figure S2.** Composition at the film surface measured by XPS collected while re-heating a decapped sample after $Ar^+$ ion sputtering to remove surface carbon and oxygen.

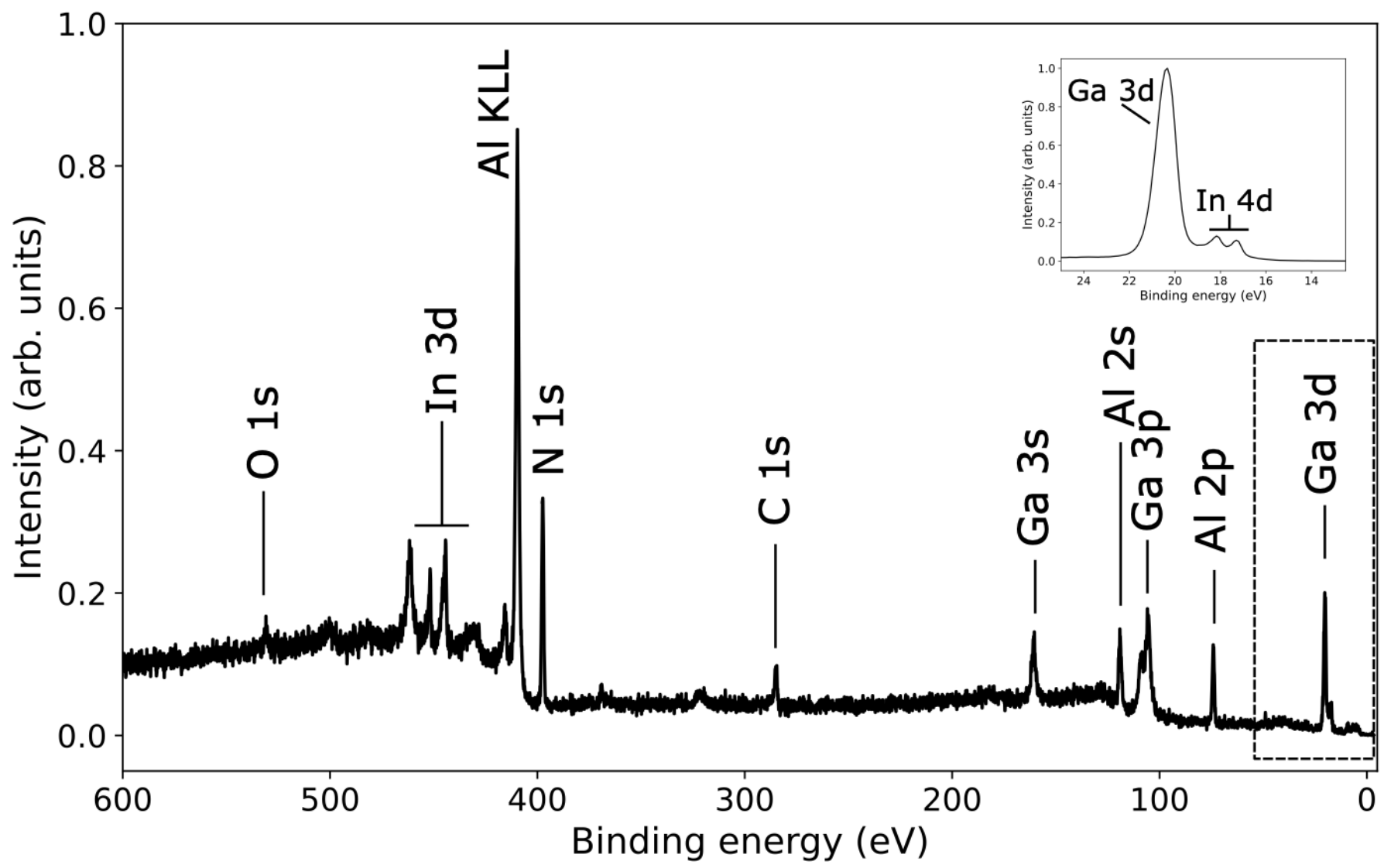

**Figure S3.** The XPS spectra taken 1800 eV after decapping *in situ* at the Advanced Photon Source at Argonne National Laboratory showing very little oxygen or carbon present at the AlGaN surface. The binding energy was calibrated by setting the adventitious carbon 1s peak to 284.8 eV. The spectra was taken at room temperature with a flood gun for charge compensation.

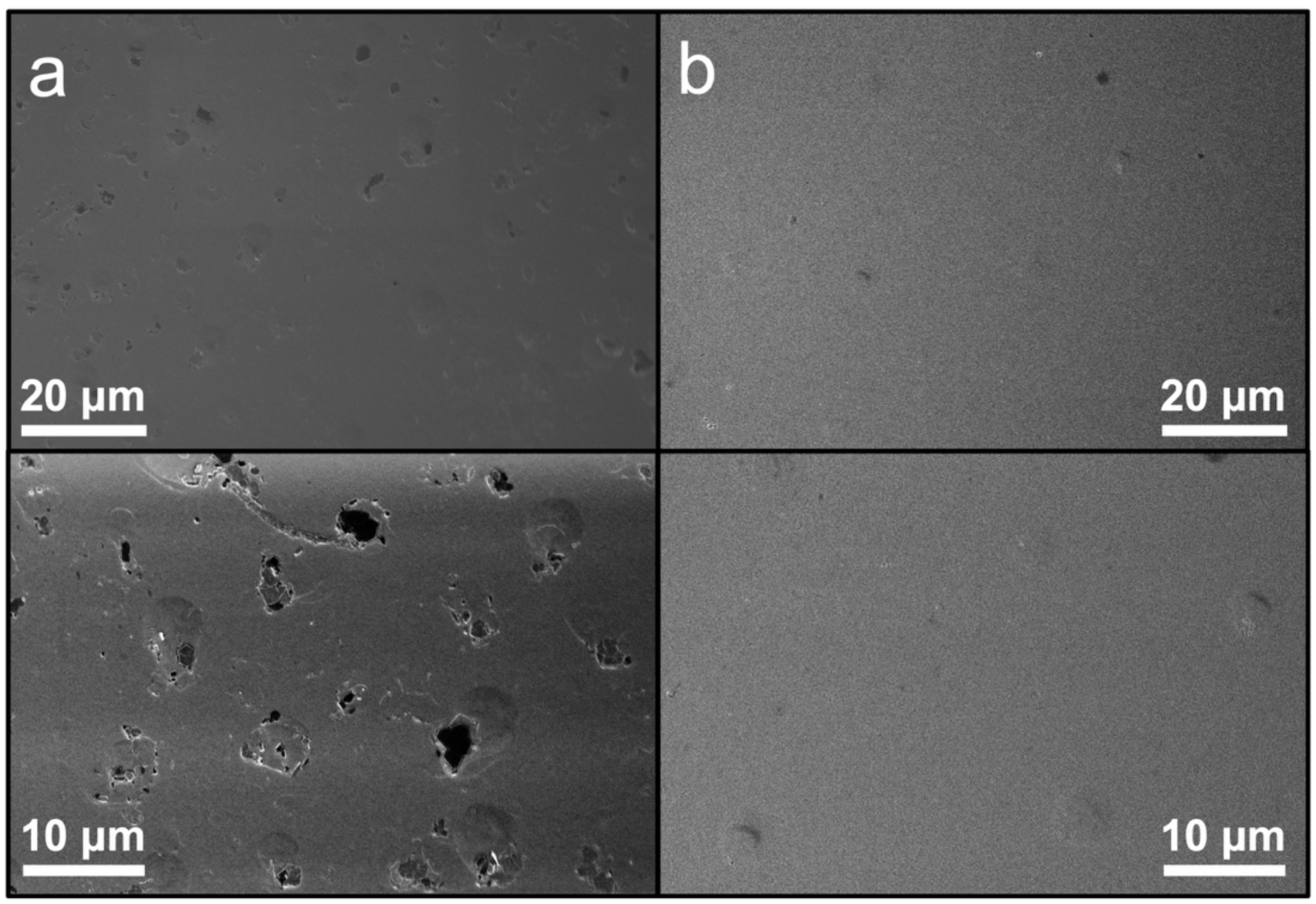

**Figure S4.** Secondary Electron SEM images of AlGaN films that were a) not capped or b) capped with InN and thermally decapped.